\documentclass[runningheads]{llncs}

\usepackage[T1]{fontenc}
\usepackage{lmodern}
\usepackage{textcomp}
\usepackage{graphicx}
\usepackage{natbib} 
\usepackage{booktabs} 
\usepackage[indentfirst=false,font={itshape},begintext=``,endtext='']{quoting}
\usepackage{subcaption}
\usepackage[hyphens,spaces,obeyspaces]{url}
\usepackage{hyperref}
\usepackage[most]{tcolorbox}
\newtcolorbox{summarybox}{
    enhanced,
    breakable,
    sharp corners,
    boxrule=0.3pt,
    colback=white,
    colframe=black!40,
    coltitle=black,
    fonttitle=\bfseries,
    toptitle=2mm,
    bottomtitle=2mm,
    colbacktitle=white,
}
\usepackage{cleveref}
\usepackage{longtable}

\urldef{\urlgamespot}\url{https://www.gamespot.com/articles/amazon-and-crytek-agree-to-licensing-deal-worth-50/1100-6426438/}

\begin{document}

\title{SyDRA: An Approach to Understand Game Engine Architecture}

\author{
Gabriel C. Ullmann\inst{1} 
\and Yann-Ga\"el Gu\'{e}h\'{e}neuc\inst{1} 
\and Fabio Petrillo\inst{2} 
\and Nicolas Anquetil\inst{3} 
\and Cristiano Politowski\inst{4} 
}

\authorrunning{G.C. Ullmann et al.}
\institute{Concordia University, Montreal, Canada \\ \email{\{gabriel.cavalheiroullmann,yann-gael.gueheneuc\}@concordia.ca} 
\and École de Technologie Supérieure, Montreal, Canada \\ \email{fabio.petrillo@etsmtl.ca} 
\and Univ. Lille, CNRS, Inria, Centrale Lille, UMR 9189 - CRIStAL \\ \email{nicolas.anquetil@inria.fr}
\and Université de Montréal, Montreal, Canada \\ \email{cristiano.politowski@umontreal.ca} }

\maketitle 

\newcommand{\saud}[1]{\textit{Audio (AUD)}}
\newcommand{\scor}[1]{\textit{Core (COR)}}
\newcommand{\sfes}[1]{\textit{Front End (FES)}}
\newcommand{\sgmp}[1]{\textit{Gameplay Foundations (GMP)}}
\newcommand{\shid}[1]{\textit{Human Interface Devices (HID)}}
\newcommand{\sllr}[1]{\textit{Low-Level Renderer (LLR)}}
\newcommand{\somp}[1]{\textit{Online Multiplayer (OMP)}}
\newcommand{\sphy}[1]{\textit{Physics (PHY)}}
\newcommand{\sdeb}[1]{\textit{Profiling \& Debugging (DEB)}}
\newcommand{\spla}[1]{\textit{Platform Independence Layer (PLA)}}
\newcommand{\sres}[1]{\textit{Resources (RES)}}
\newcommand{\sska}[1]{\textit{Skeletal Animation (SKA)}}
\newcommand{\ssgc}[1]{\textit{Scene Graph / Culling Optimizations (SGC)}}
\newcommand{\ssdk}[1]{\textit{Third-Party SDKs (SDK)}}
\newcommand{\svfx}[1]{\textit{Visual Effects (VFX)}}
\newcommand{\sedi}[1]{\textit{World Editor (EDI)}}

\vspace{-0.8cm}
\begin{abstract}
Game engines are tools to facilitate video game development. They provide graphics, sound, and physics simulation features, which would have to be otherwise implemented by developers. Even though essential for modern commercial video game development, game engines are complex and developers often struggle to understand their architecture, leading to maintainability and evolution issues that negatively affect video game productions. In this paper, we present the Subsystem-Dependency Recovery Approach (SyDRA), which helps game engine developers understand game engine architecture and therefore make informed game engine development choices. By applying this approach to 10 open-source game engines, we obtain architectural models that can be used to compare game engine architectures and identify and solve issues of excessive coupling and folder nesting. Through a controlled experiment, we show that the inspection of the architectural models derived from SyDRA enables developers to complete tasks related to architectural understanding and impact analysis in less time and with higher correctness than without these models.
\end{abstract}


\section{Introduction}
\label{sec:introduction}
Game engines are tools to facilitate and accelerate video game development. They provide out-of-the-box features that are broad enough to be used to create a variety of video games, for example, graphics rendering, sound management, and physics simulation. Developing a game engine from scratch is expensive and time-consuming, an endeavour only large video game companies can afford. While risky, such endeavour brings benefits such as the possibility of optimizing the tool for a specific kind of game or game genre. Moreover, by creating their in-house solution, companies avoid paying licensing fees while being free to license their technology to others, such as CryEngine\footnote{\urlgamespot} and IdTech\footnote{\url{https://www.eurogamer.net/id-tech-5-only-for-bethesda-titles}} have done recently.

However, the advantages of developing a game engine only last as long as the understanding of its structure remains. Over time, development teams change due to high turnover \citep[p.~292]{cadin:halshs-00271887}, and the game engine code base changes, which causes it to drift away from its original structure, in a process known as architectural drift or erosion \cite[p.~2]{baabad_software_2020}. Moreover, due to the need to innovate and experiment with different video games, companies sometimes decide to re-purpose their game engines by developing new features on top of a legacy architecture. In this scenario, the lack of understanding of the original architecture of the core game engine code causes maintenance and evolution problems. An example of this kind of problem was reported by developers of the Frostbite game engine, used by Bioware to develop Anthem in 2019\footnote{\url{https://kotaku.com/how-biowares-anthem-went-wrong-1833731964}}:

\begin{quotation}
\noindent
``Frostbite is like an in-house engine with all the problems that entails—it's poorly documented, hacked together, and so on—with all the problems of an externally sourced engine,'' said one former BioWare employee. ``Nobody you actually work with designed it, so you don't know why this thing works the way it does, why this is named the way it is.''
\end{quotation}

The lack of understanding of the architecture of a game engine hinders a developer's capacity to maintain and evolve it, considerably affecting the schedule of a game development project. For example, the re-purposing of Frostbite to Dragon Age: Inquisition ``took up about a third of the project's development time''\footnote{\url{https://gamerant.com/dragon-age-mark-darrah-bioware-problems/}}. Similarly, developers from Bethesda added multiplayer support to Creation Engine, originally made for single-player games only. During the development of Fallout 76, this feature caused several bugs and ``put additional time pressure on the schedule''\footnote{\url{https://kotaku.com/bethesda-zenimax-fallout-76-crunch-development-1849033233}}. Therefore, we can claim architectural understanding impacts both the technical and managerial sides of video game development.

While the problems reported by Frostbite and Creation Engine developers reflect the reality of closed-source game engine development, similar problems can be observed in open-source game engines. Even though game engines such as Unreal and Godot maintain official documentation and support forums, these data sources mostly focus on providing video game developers with a way to ``get started''\footnote{\url{https://dev.epicgames.com/documentation/en-us/unreal-engine/understanding-the-basics-of-unreal-engine}} by explaining how to use the game engine's features, while hiding low-level aspects of their implementation. Therefore, game engine developers wishing to re-purpose a game engine or choose which best fits their needs cannot rely solely on this documentation. They must plunge into the code, study its structures, and finally compare them with those of other game engines to make informed game engine development choices. 

In this paper, we use the Subsystem Dependency Recovery Approach (SyDRA), described in our previous work \citep{ciancarini_visualising_2023}, to help game engine developers understand game engine architecture and therefore make informed game engine development choices. By applying SyDRA to 10 open-source game engines, we obtain architectural models that can be used to compare game engine architectures and identify and solve problems such as high coupling and low cohesion. Through a controlled experiment with 16 participants, we show that the inspection of the architectural models derived from SyDRA enables developers to complete tasks related to architectural understanding and impact analysis in less time and with higher correctness than without these models.

The implementation and evaluation of SyDRA enable us to answer the following research questions:
\begin{itemize}
    \item \textbf{RQ1}: Does SyDRA help developers \textit{understand} game engines more effectively than by simply inspecting the code?
    \item \textbf{RQ2}: Does SyDRA help game engine developers perform game engine \textit{impact analysis} more effectively than by simply inspecting the code?
\end{itemize}

We answer positively to these two questions and conclude that by using SyDRA, developers can better understand game engine architecture without increasing their perceived workload. Our experiments also provide insights into how developers perceive workload in software analysis tools and how it correlates with their professional experience. 

The paper is organised as follows. \Cref{sec:related-work} presents related work on software architecture and game engines. \Cref{sec:approach} provides a high-level description of SyDRA, our software architecture recovery approach, and how it was used to generate an architectural model of 10 open-source game engines. \Cref{sec:results} describes how the architectural models resulting from SyDRA can help game engine developers solve folder organization and coupling issues, as well as make game engine development choices. \Cref{sec:evaluation} describes the design and execution of a user study and its results. \Cref{sec:threats} presents internal and external threats to validity, and \Cref{sec:conclusion} presents the conclusion and future work. 

\section{Related Work}
\label{sec:related-work}
The structure and purpose of game engines are often not well-documented, and for this reason, several researchers have attempted to use software architecture recovery techniques to search for architectural structures in the source code and help developers assign them meaning. For example, \citeauthor{munro_architectural_2009} (\citeyear[p.~247]{munro_architectural_2009}) used Doxygen\footnote{\url{https://www.doxygen.nl}}, a popular documentation generation tool, to extract dependency information from an open-source version of the IdTech game engine. This data was then used to create dependency graphs, which aided ``in the process of identifying suitable improvements and enhancements to a specific engine and have supported implementing these in an appropriate manner''.

\citeauthor{agrahari_whats_2021} (\citeyear[p.~1]{agrahari_whats_2021}) developed and used AC$^2$, a software analysis tool ``to generate call graphs and collaboration graphs across three releases'' of Unreal Engine. These graphs helped them identify architectural patterns in components, observe their evolution and ``aid in better comprehension of this complex and widely used game engine for researchers and practitioners''.

Researchers conduct experiments to evaluate the usefulness of extracted architectural models, for example, \cite{heijstek_experimental_2011} and \cite{briand_controlled_2001}. In these studies, they present a set of architectural understanding tasks to developers. By measuring how swiftly and correctly developers can perform tasks with and without a supporting architectural model or documentation, they assess the benefit this model brings to system understanding. In \Cref{sec:evaluation}, we describe how we designed and conducted a controlled experiment based on the work of \cite{briand_controlled_2001}.

Alternatively, researchers may conduct field studies, observing how developers and architects use them in real scenarios \citep[p.~1]{abi-antoun_field_2008}. However, this kind of study requires collaboration with companies and long-term observation. Considering the closed-source nature of video game and game engine development, this would not be a viable option for our study. As we explain in \Cref{sec:approach}, SyDRA relies on the analysis of source code and documentation, both of which must be openly available.
\vspace{-0.35cm}
\section{Approach}
\label{sec:approach}
\begin{figure*}[ht]
  \center
  \includegraphics[width=0.87\linewidth]{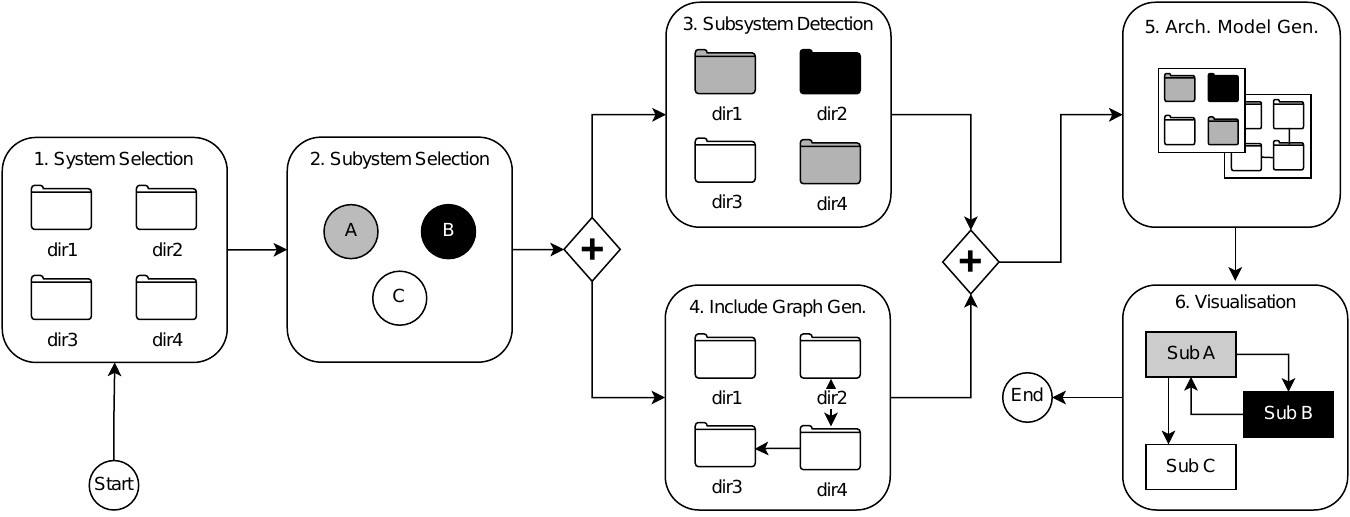}
  \caption{Steps of our game engine analysis approach.}
  \label{fig:diagram-approach}
\end{figure*}

\vspace{-0.25cm}
The Subsystem Dependency Recovery Approach (SyDRA) comprises six steps, shown in \Cref{fig:diagram-approach}, and it helps game engine developers to analyse one or more game engines. SyDRA's steps describe how a developer can consistently select game engines to analyse, cluster their files and folders and later cross-referencing this information with an \textit{include} graph. As a result, they obtain architectural models that can be used to understand a game engine's architectural structure, identify and reduce excessive coupling, and increase cohesion.

Our implementation of SyDRA is available on GitHub\footnote{\url{https://github.com/gamedev-studies/game-engine-analyser}}, along with data and visualisations resulting from our analysis of 10 open-source game engines, which we also show and discuss in previous work \citep{ciancarini_visualising_2023}. We obtain these results by implementing SyDRA's steps as follows:

\textbf{1. System Selection}: We select 10 open-source game engines from GitHub, searching for the ``game engine'' keyword, filtering by the language C++, and then ordering the results by their popularity. We consider popularity to be the number of GitHub stars in a repository. We chose this metric because it is ``partial evidence for the repository containing an engineered software project'' \cite[p.7]{munaiah_curating_2017}. This way, we avoid selecting ``toy'' projects which do not properly represent the scale and complexity of industry-grade game engines.

\textbf{2. Subsystem Selection}: For each selected game engine, we define how to cluster each file and folder into functional groups, which are subsystems. We consider 16 subsystems described in the ``Runtime Game Engine Architecture'' by \cite[p.~33]{gregory_game_2018}, each corresponding to common features needed to create most video games. For example, graphics, audio, physics simulation and input device processing. We summarize this list of subsystems as shown in \Cref{tab:gregory-summarized}.

\vspace{0.5cm}
\begin{longtable}{p{.1\textwidth} p{.3\textwidth} p{.5\textwidth}} 
\hline
\textbf{ID} & \textbf{Subsystem} & \textbf{Description} \\ \hline

AUD                 & Audio                             & Manages audio playback and effects.\newline                                              \\ \hline
COR                 & Core                       & Manages engine initialisation and contains libraries for math, memory allocation, etc.  \\ \hline
DEB                 & Profiling \& Debugging           & Manages performance stats, debugging via in-game menus or console.                   \\ \hline
EDI                 & World Editor                      & Enables visual game world-building.                                   \\ \hline
FES                 & Front End                         & Manages GUI, menus, heads-up display (HUD), and video playback.                      \\ \hline
GMP                 & Gameplay Foundations              & Manages the game object model, scripting and event/messaging system.                 \\ \hline
HID                 & Human Interface \phantom{aaaaa}\linebreak{} Devices           & Manages game-specific input interfaces, physical I/O devices.                        \\ \hline
LLR                 & Low-Level Renderer                & Manages cameras, textures, shaders, fonts, and general drawing tasks.                \\ \hline
OMP                 & Online Multiplayer                & Manages match-making and game state replication.                                     \\ \hline
PHY                 & Collision \& Physics             & Manages forces and constraints, rigid bodies, ray/shape casting.                     \\ \hline
PLA                 & Platform Independence Layer       & Manages platform-specific graphics, file systems, threading, etc.                    \\ \hline
RES                 & Resources                         & Manages the loading/caching of game assets, such as 3D models, textures, fonts, etc. \\ \hline
SDK                 & Third-Party SDKs                  & Enables interfacing with DirectX, OpenGL,  Havok, PhysX, STL, etc.            \\ \hline
SKA                 & Skeletal Animation                & Manages animation state tree, inverse kinematics (IK), and mesh rendering.           \\ \hline
SGC                 & Scene Graph/ \phantom{aaaaaaaa}\linebreak{} Culling Optimizations & Computes spatial hash, occlusion, and level of detail (LOD).                         \\ \hline
VFX                 & Visual Effects                    & Enables light mapping, dynamic shadows, particles, decals, etc.                      \\ \hline
\caption{Summarized ``Runtime Game Engine Architecture'' subsystem descriptions, adapted from \citeauthor{gregory_game_2018} (\citeyear[p.~33]{gregory_game_2018})}
\label{tab:gregory-summarized}
\end{longtable}

\textbf{3. Subsystem Detection}: For each selected game engine, we manually cluster all files and folders into the selected subsystems. To ensure files implementing the same features are clustered into the same group, we consider their naming, folder hierarchy, mentions in the documentation and the comments found in their source code to determine their functionality.

\textbf{4. Include Graph Generation}: For each selected game engine, we generate an \textit{include} graph which represents dependencies between files. This step is done automatically with the \textit{cinclude2dot} tool\footnote{\url{https://www.flourish.org/cinclude2dot/}}.

\textbf{5. Architectural Model Generation}: For each selected game engine, we use the data obtained from Steps 3 and 4 to generate an architectural model. We then load these models into Moose, a software analysis platform which we describe in more detail in \Cref{sub:exp-materials}. This step is done semi-automatically.

\textbf{6. Architectural Model Visualisation}: For each selected game engine, we use the visualisation titled ``Architectural map'' from Moose to generate a visual representation of the \textit{include} graph, files, folders and subsystems. This step is done semi-automatically.

\begin{figure}[ht]
    \centering
    \begin{subfigure}{0.25\linewidth}
        \centering
        \includegraphics[width=\linewidth]{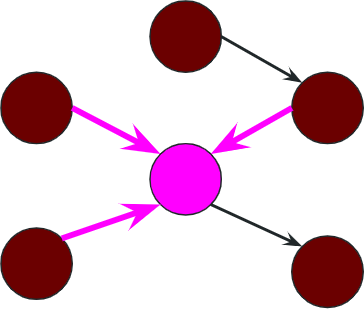}
        \caption{In-degree}
        \label{subfig:in-degree}
    \end{subfigure}
    \hfill
    \begin{subfigure}{0.25\linewidth}
        \centering
        \includegraphics[width=\linewidth]{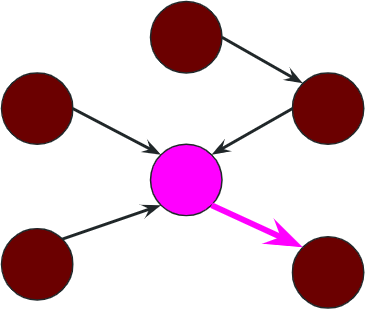}
        \caption{Out-degree}
        \label{subfig:out-degree}
    \end{subfigure}
        \hfill
    \begin{subfigure}{0.25\linewidth}
        \centering
        \includegraphics[width=\linewidth]{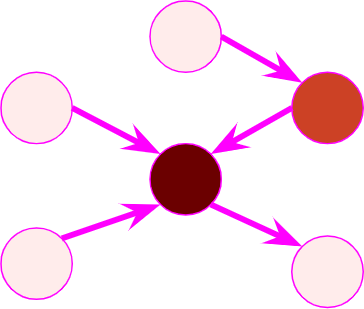}
        \caption{Centrality}
        \label{subfig:centrality}
    \end{subfigure}
    \caption{ Graph metrics used in our analysis of SyDRA's results }
     \label{fig:arch-map-exs}
\end{figure}

Additionally, based on the number of nodes and edges on each \textit{include} graph, we compute the following metrics, illustrated in \Cref{fig:arch-map-exs}: 
\begin{itemize}
    \item \textbf{In-degree}: The count of incoming edges of a node (e.g., the Physics subsystem is included by three other subsystems).
    \item \textbf{Out-degree}: The count of outgoing edges of a node (e.g., the Physics subsystem includes one other subsystem).
    \item \textbf{Betweenness centrality}: The extent to which a node lies in the path of others \cite[p.~758]{badar_examining_2013} (e.g., most subsystems include the Physics subsystem, which in turn includes others).
\end{itemize}

It took us approximately six months to apply SyDRA's steps to 10 open-source game engines. The most time-consuming step was subsystem detection (Step 3), performed exclusively by the first author over approximately two months. In \Cref{sec:results}, we present the architectural models derived from the application of SyDRA. Furthermore, in \Cref{sec:evaluation}, we show examples of how the use of architectural models can inform developers performing architecture understanding and impact analysis tasks.
\vspace{-0.2cm}
\section{Application}
\label{sec:results}
\begin{figure}[h]
    \centering
    \begin{subfigure}{0.45\linewidth}
        \centering
        \includegraphics[width=\linewidth]{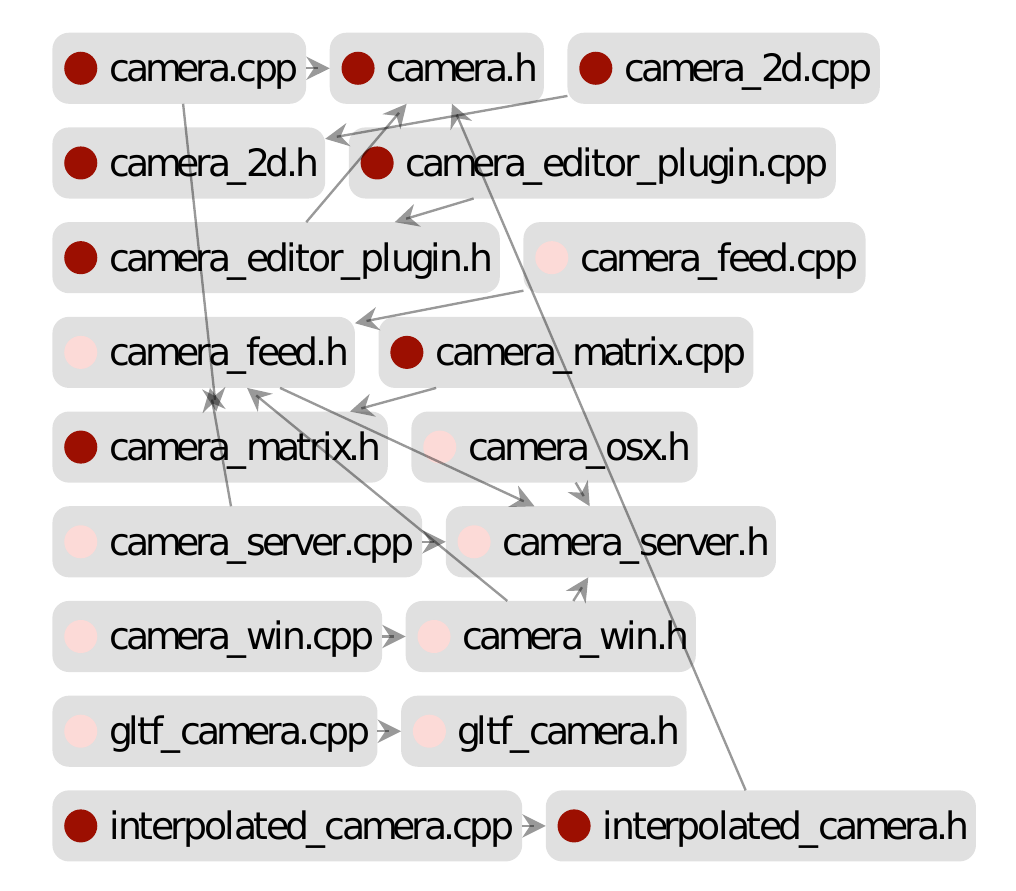}
        \caption{Viewing unclustered files}
        \label{subfig:view-rel-files}
    \end{subfigure}
    \hfill
    \begin{subfigure}{0.5\linewidth}
        \centering
        \includegraphics[width=\linewidth]{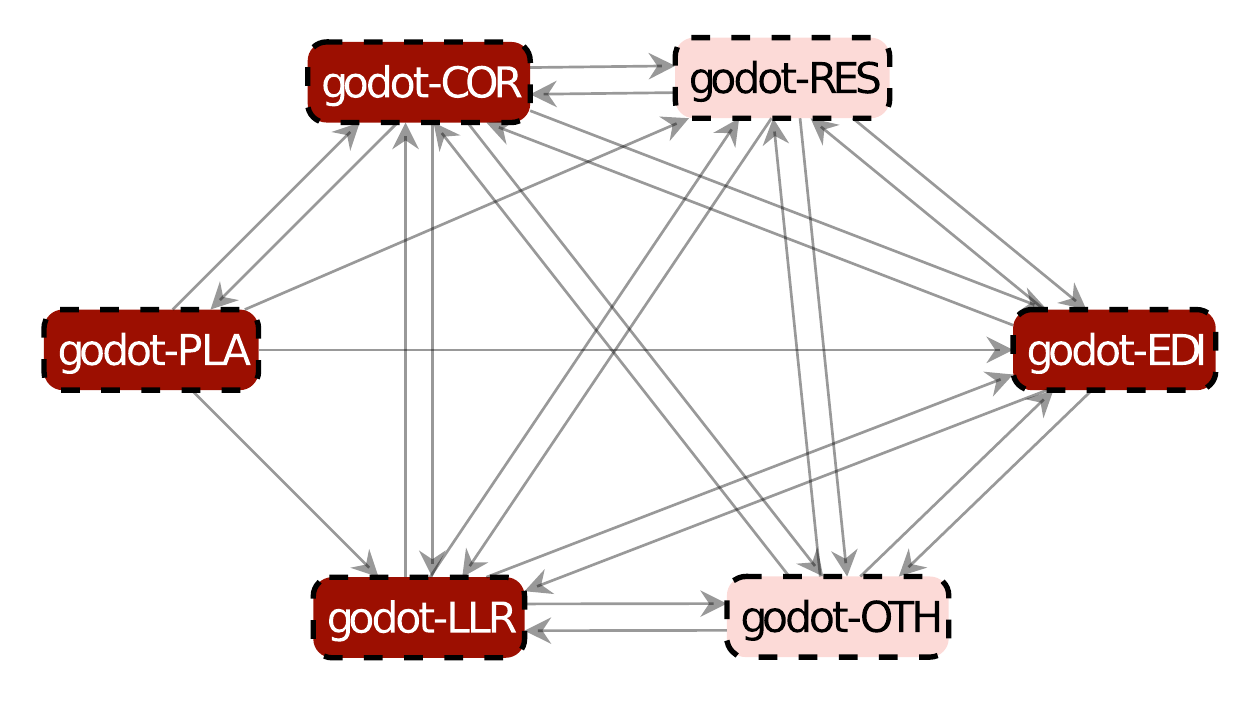}
        \caption{Viewing files clustered by subsystem}
        \label{subfig:view-rel-subs}
    \end{subfigure}
    \caption{Architectural Map showing files containing the word ``camera'' from Godot }
     \label{fig:arch-map-clustering}
\end{figure}

As a result of applying the SyDRA, we obtained 10 architectural models, one for each of the selected game engines. These models are graphs where each node is a file, and each edge is an \textit{include} relationship between files. For example, using the Architectural Map from Moose, these files can be viewed clustered by subsystem, or individually, as shown in \Cref{fig:arch-map-clustering}. 

Examining subsystems and their relationships in the Architectural Map serves as a starting point for game engine developers seeking to understand the structure and functionality of a game engine. In this section, we show some examples of visual and numerical information extracted from these 10 architectural models and how they help developers understand three aspects of game engines: subsystem coupling, subsystem cohesion and coupling between files and subsystems.

\subsection{Subsystem Coupling}
\label{sec:subsystem-aspect}
\begin{figure}[ht]
    \centering
    \includegraphics[width=\linewidth]{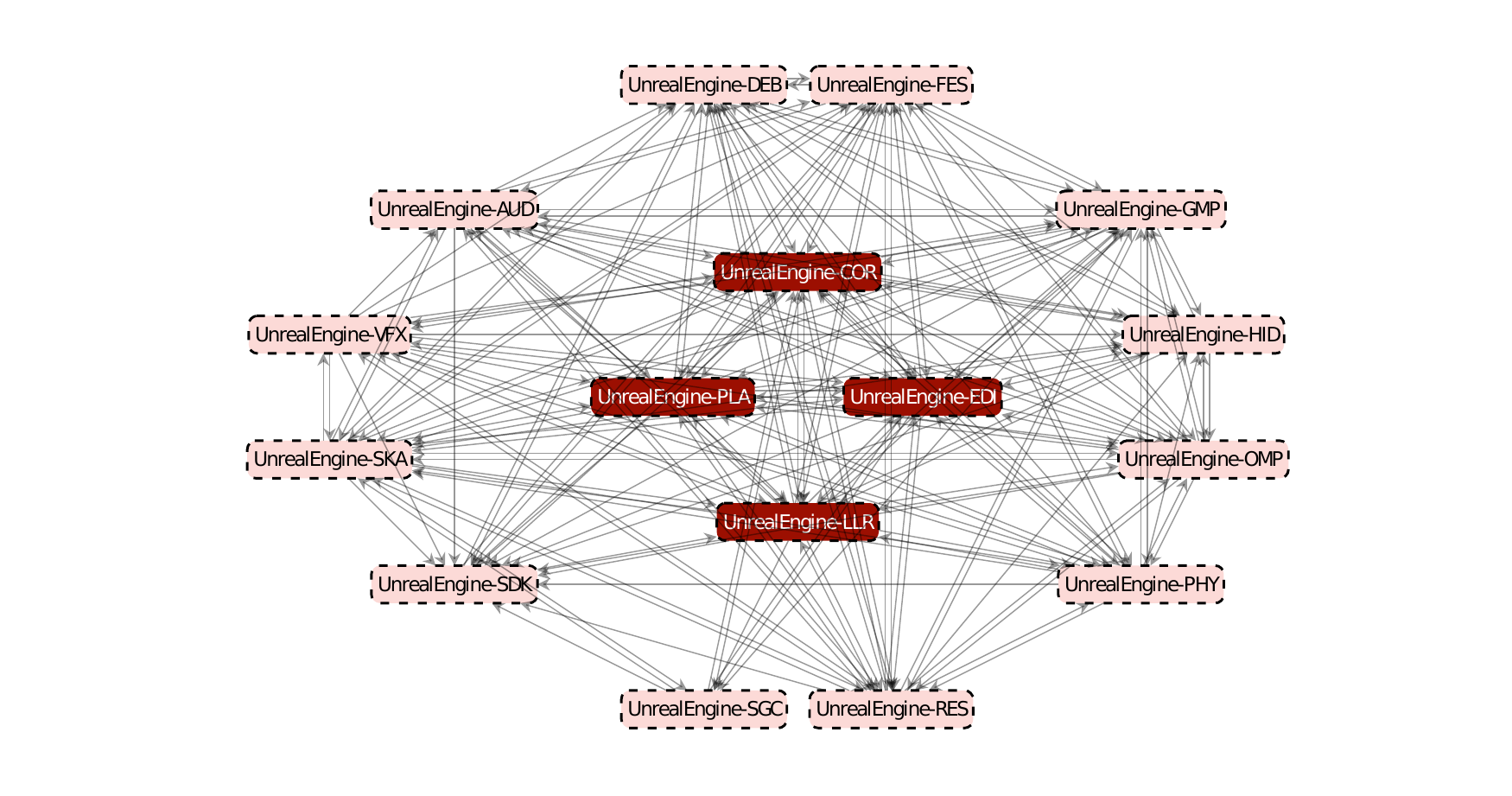}
    \caption{Unreal Engine's architectural model.}
    \label{fig:unreal_arch_model}
\end{figure}

The models produced by the SyDRA enable us to visualize the high-level architecture of the game engine, meaning it shows us what features are available and how they depend on each other. In Unreal Engine and Godot, we detected all 16 subsystems described in the reference architecture (see \Cref{fig:unreal_arch_model}). In the remaining game engines, we detected 12 or more subsystems. The only exception was OlcPixelGameEngine, in which we detected only five subsystems. Therefore, 90\% of the game engines we analysed contain at least 75\% of the subsystems described in the reference architecture, which shows there are many similarities between the reference architecture and the actual architecture of open-source game engines.

The \scor{}, \sllr{} and \sres{} are the most frequently coupled subsystems, as shown in \Cref{tab:results_frequent_coupling}. As shown in \Cref{tab:gregory-summarized}, these subsystems have several responsibilities, and we believe this is the reason behind their high coupling. For example, the \sllr{} subsystem goes beyond simply drawing on the screen and also provides abstractions that, while visual, also relates to other subsystems. For example, while camera functionality is part of \sllr{}, to know what is within the view of the camera and therefore, what needs to be drawn, it depends on information from the \ssgc{} subsystem, which provides culling and occlusion computation.

\scor{}, \sllr{} and \sres{}  have high in-degree and centrality, along with \spla{}, which likewise centralizes utilities and cross-platform compatibility code. While this type of analysis enables us to detect coupling patterns, it cannot explain why this coupling exists and whether it could be reduced. For this reason, in \Cref{sec:coupling-aspect}, we show examples that explain why certain subsystems are more coupled in certain game engines, whether these coupling patterns repeat in different game engines and what they show about the architecture.

\begin{table}[ht]
\centering
\caption{The top frequent subsystem coupling pairs.}
\label{tab:results_frequent_coupling}
\begin{tabular}{lclr|lclr}
\hline
\textbf{Pair} & \textbf{} & \textbf{} & \textbf{Count\phantom{a}} & \textbf{\phantom{a}Pair} & \textbf{} & \textbf{} & \textbf{\phantom{a}Count} \\ \hline
COR & $\rightarrow{}$ & LLR & 9\phantom{a}     & \phantom{a}COR & $\rightarrow{}$ & PHY & 8 \\
GMP & $\rightarrow{}$ & COR & 9\phantom{a}     & \phantom{a}FES & $\rightarrow{}$ & COR & 8 \\
LLR & $\rightarrow{}$ & COR & 9\phantom{a}     & \phantom{a}RES & $\rightarrow{}$ & COR & 8 \\
PHY & $\rightarrow{}$ & COR & 9\phantom{a}     & \phantom{a}SKA & $\rightarrow{}$ & COR & 8 \\ 
COR & $\rightarrow{}$ & RES & 8\phantom{a}     & \phantom{a}SKA & $\rightarrow{}$ & LLR & 8 \\ 
LLR & $\rightarrow{}$ & RES & 8\phantom{a}     &     &                 &     &   \\ \bottomrule
\end{tabular}
\end{table}

\begin{figure}[ht]
    \centering
    \includegraphics[width=0.9\textwidth]{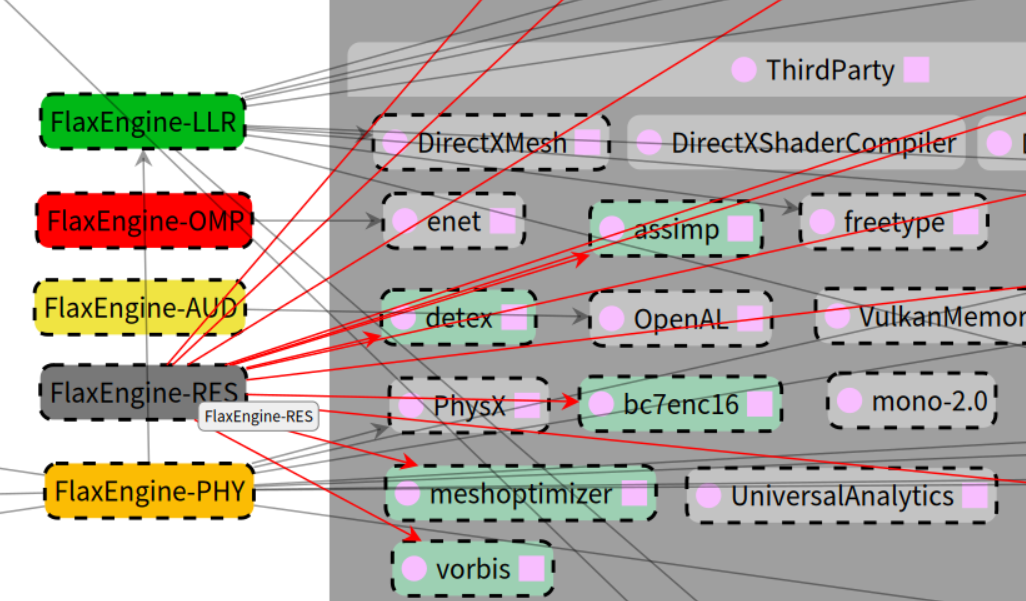}
    \caption{FlaxEngine \textit{include} relationships between subsystems and \ssdk{}}
    \label{fig:flaxengine_thirdparty}
\end{figure}

While architectural models can give us insights into groups of subsystems, or even groups of game engines, we can also focus on a particular subsystem of a particular game engine to understand its features. For example, on \Cref{fig:flaxengine_thirdparty}, we show that, upon inspection of the \ssdk{} subsystem of FlaxEngine, we can view the \textit{include} relationship between its libraries and several subsystems, and from this relationships, we can infer the functionality of the libraries, even if we do not know their functionality. For example, we can infer \textit{DirectXMesh} is a graphics-related library, due to it being included by with the \sllr{} subsystem. Similarly, \textit{detex}\footnote{\url{https://github.com/hglm/detex}}, a texture decompression library, is included by the \sres{} subsystem, which is responsible for file loading and management.

In addition to their functional features, understanding the availability of specific libraries within the \ssdk{} can offer valuable insights into both architectural considerations and functional limitations. For example, Unreal Engine uses PhysX, a physics library optimized to leverage GPU acceleration typically found in NVidia video cards. Conversely, Godot uses Bullet, and cannot benefit from this hardware performance boost. Consequently, when game engine developers are faced with a decision between two platforms based on their physics capabilities, this insight can serve to inform their choice. While present in the code and folder structures, the inspection of the architectural model helps make this information more evident to the game engine developer.

\subsection{Subsystem Cohesion}
\label{sec:folder-level-ref}

Especially on game engines that have been in development for decades, such as Unreal, Godot and O3DE, the folder structure tends to grow large and sometimes become excessively nested. Also, a small number of folders contains the majority of files, which is an indicator of low file cohesion. Architectural models based on files and folders, such as those produced by SyDRA, highlight where the bulk of the source code is located within the directory structure, and therefore points of low file cohesion in this structure. For example, in O3DE, most of the files implementing subsystem features are located in \textit{./Code/Framework}. By analysing the subfolder organisation of each folder we observe that the folders \textit{AzCore}, \textit{AzFramework}, \textit{AzNetworking} and \textit{AzToolsFramework} share the same folder organisation pattern, comprised as follows:
\begin{itemize}
    \item \textbf{Features:} Contain files that implement features. This folder always has the same name as its parent (e.g., \textit{AzCore} has a folder also named \textit{AzCore}).
    \item \textbf{Platform:} Contains files that implement a part of \spla{} subsystem related to the ``features'' folder.
    \item \textbf{Tests:} Contains unit tests to features in the ``features'' folder.
\end{itemize}

\begin{figure}[ht]
    \centering
    \includegraphics[width=0.9\textwidth]{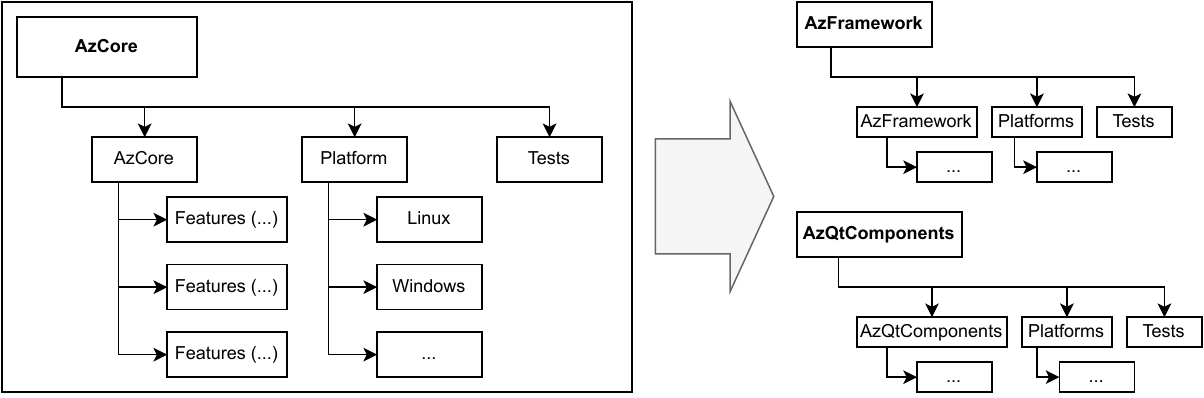}
    \caption{Folder organisation pattern we found in O3DE}
    \label{fig:o3de-actual}
\end{figure}

We demonstrate this folder organisation pattern in \Cref{fig:o3de-actual}. While its rationale is not explained in O3DE's documentation, we believe it was created to break down the \spla{}, a large subsystem that encompasses many features, into folders with a lower file count. However, while highlighting the separation between O3DE's core and other subsystems, this organisation does not use its folder hierarchy to cohesively group all subsystem code under one folder. For example,\textit{AzQtComponents}, which is a part of the \sfes{} subsystem, is in the same hierarchy level as \textit{AzFramework}, a folder which contains files for several subsystems.

\begin{figure}[ht]
    \centering
    \includegraphics[width=0.55\textwidth]{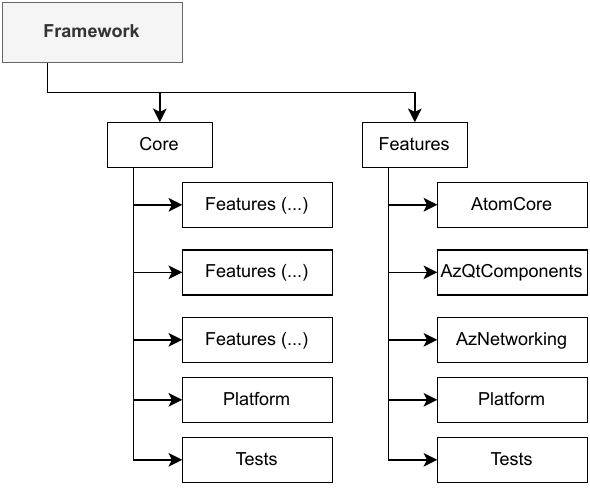}
    \caption{Alternative folder organisation for O3DE}
    \label{fig:o3de-suggestion}
\end{figure}

In \Cref{fig:o3de-suggestion}, we show an alternative organisation that has two folders in its top level, ``Core'' and ``Features'', which are then subdivided by purpose: containing code from O3DE's core only or from other subsystems. Each of these top-level folders keeps its own \textit{Platform} and \textit{Test} folders. Thus, we avoid naming repetition and create a more semantic folder hierarchy, which separates the subsystems and their features from the higher-level concept of ``Core'' vs. ``Features''.

While at first glance such folder-level refactoring may appear merely aesthetic, it can help developers re-purpose game engines. By eliminating unnecessary nesting and clustering files by subsystem, developers can focus on going forward with the development of new features without being encumbered by legacy architectural structures and naming that no longer serve a purpose.

\subsection{File and Subsystem Coupling}
\label{sec:coupling-aspect}

\begin{figure}[ht]
  \begin{subfigure}{0.48\linewidth}
      \centering
      \includegraphics[width=\linewidth]{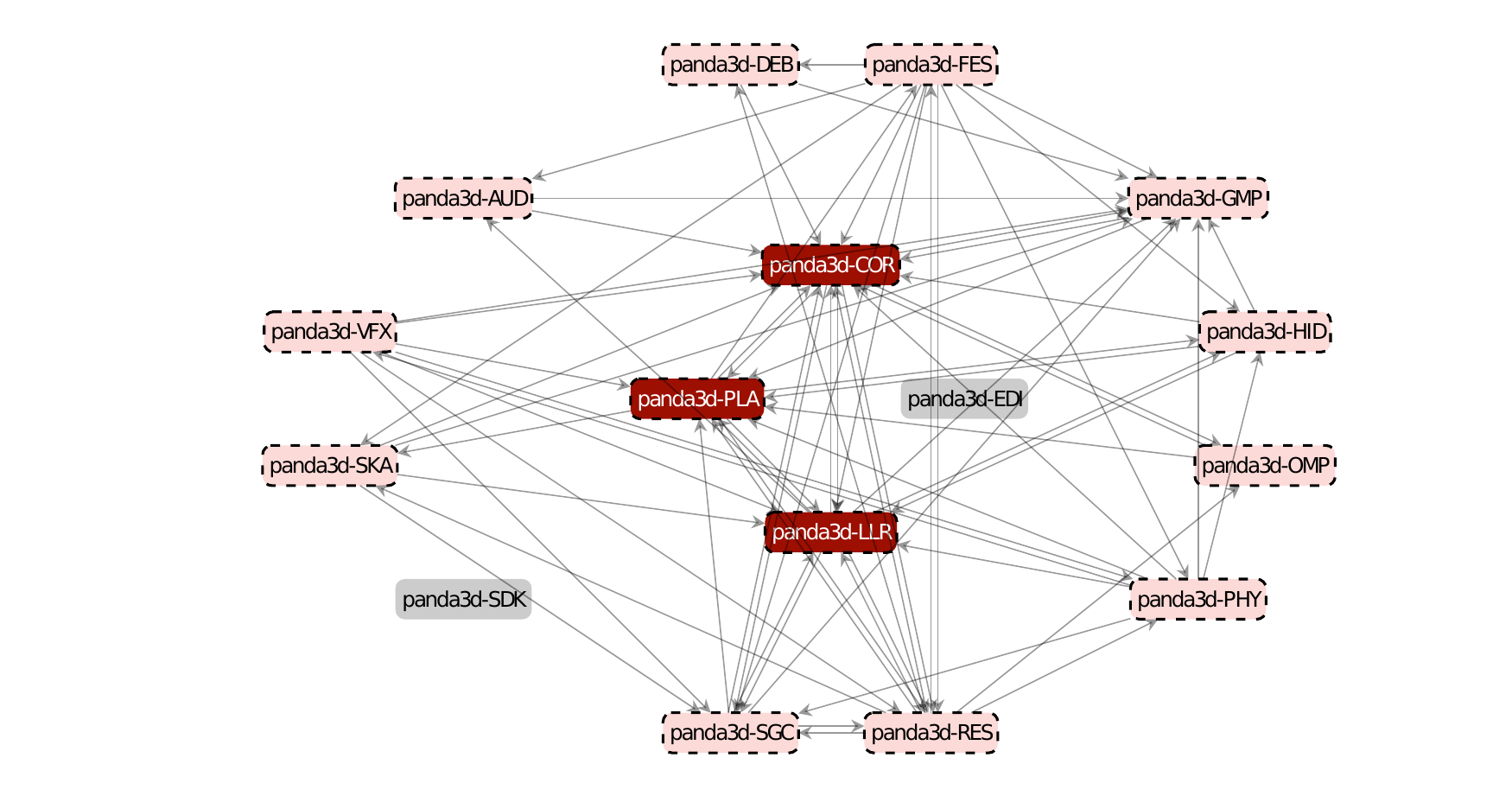}
      \caption{Panda3D}
      \label{fig:panda3d_arch_model}
  \end{subfigure}
  \begin{subfigure}{0.48\linewidth}
      \includegraphics[width=\linewidth]{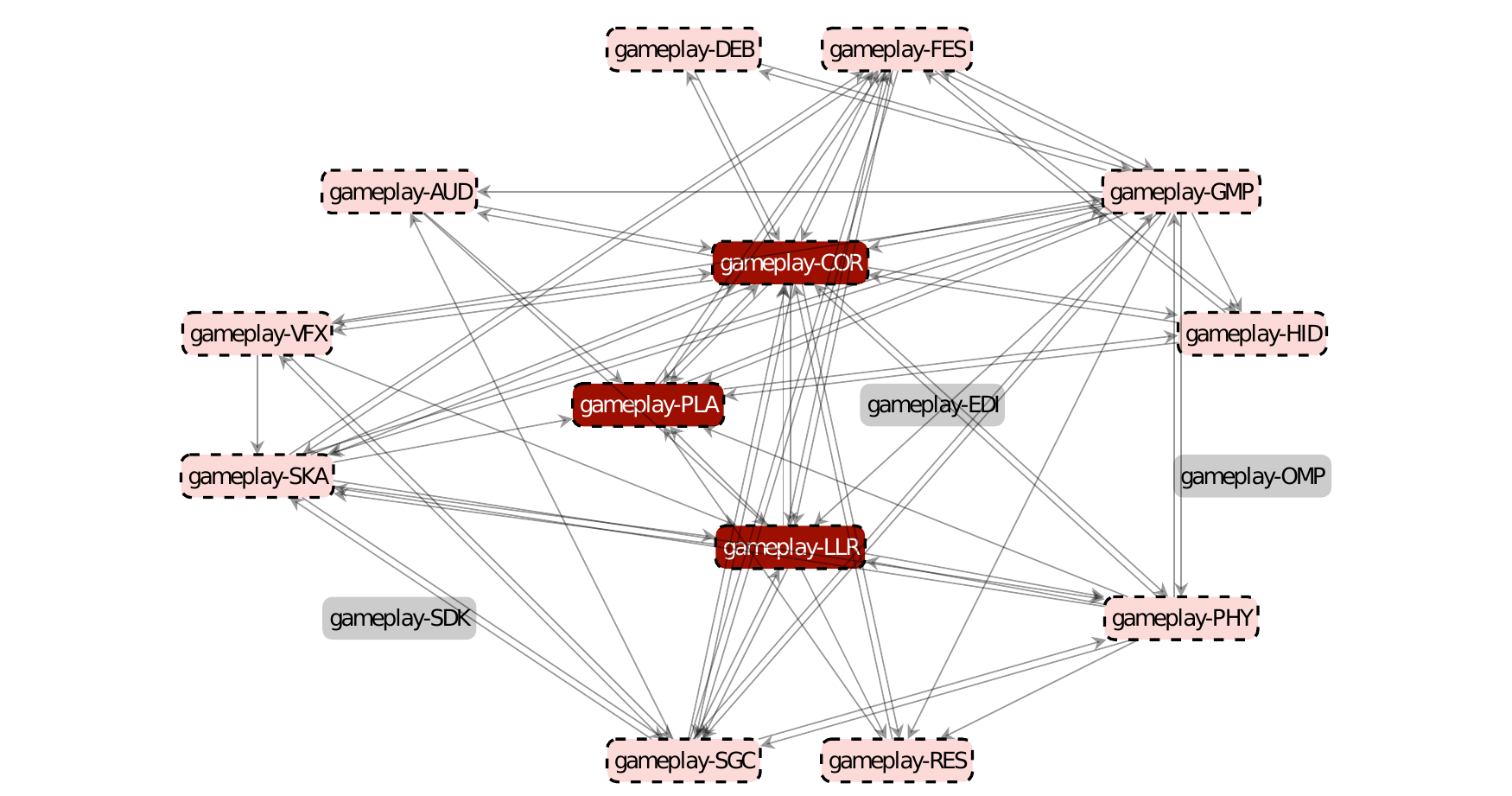}
      \caption{GamePlay3d}
      \label{fig:gameplay3d_arch_model}
  \end{subfigure}
  \caption{ Architectural models }
  \label{fig:file_and_sub_coupling}
\end{figure}

The inspection of the \textit{include} relationships between files and the cross-referencing of this information with game engine documentation, when it exists, may give us detailed insights into how a particular part of a subsystem works. In this section, we inspect subsystems in Panda3D and GamePlay3d, as shown in \Cref{fig:file_and_sub_coupling}.

For example, if we look into Panda3D's \sllr{} subsystem, we observe it has a high in-degree because of its \textit{display} folder, which contains files implementing graphics-related functionality used by the \ssgc{} and \svfx{} subsystems. For instance, one of these files, \textit{graphicsStateGuardian.h}, implements a graphics state guardian (GSG), which receives high-level rendering instructions (e.g., drawing a character present in the scene graph) and then handles low-level rendering instructions in a format the operating system and graphics hardware can understand. As explained by \citeauthor{goslin_panda3d_2004} (\citeyear[p.~112]{goslin_panda3d_2004}):

\begin{quotation}
\noindent
``All code specific to rendering on a particular platform is contained within a well-defined class called a graphics state guardian. After the system transforms and culls the scene graph, it hands off the graphics entities to the GSG for rendering. A game or application only needs to interact with the scene graph, which means the only part of the code that the system must port and optimize for a particular hardware platform is the local version of the GSG class itself.''
\end{quotation}

In GamePlay3d, the \sdeb{} subsystem has a high in-degree because its \textit{DebugNew.h} file is included by several subsystems to replace global \textit{new} and \textit{delete} C++ operators for ``memory tracking''\footnote{\url{https://github.com/gameplay3d/gameplay/blob/4de92c4c6/gameplay/src/DebugNew.h}}. We observe a similar implementation in Urho3d. We observe the \textit{Logger.h} file is also frequently included for debugging purposes. Even though debugging code would normally be removed upon pushing to the \textit{master} branch, we observe four files in GamePlay3d's repository still include either \textit{DebugNew.h} or \textit{Logger.h}. 

These are just two examples of how detailed information about subsystem functionality can help game engine developers understand the subsystems they are working with. This information is essential during the planning of game engine re-purposing, so game developers can be aware of the existing architectural structures and how the change or removal of one impacts others. In \Cref{sec:evaluation}, we show other examples of how the use of architectural models can support architectural understanding and inform developers on impact analysis.

\subsection{Discussion}
\label{sec:discussion}

\fbox{
    \parbox{0.95\textwidth}{
        "And you see every time I made a further division, up came more boxes based on these divisions until I had a huge pyramid of boxes. Finally you see that while I was splitting the cycle up into finer and finer pieces, I was also building a structure. (...) The overall name of these interrelated structures, the genus of which the hierarchy of containment and structure of causation are just species, is a system." \\
        \textbf{Robert M. Pirsig, Zen and the Art of Motorcycle Maintenance}
    }\vspace{1cm}
}
\vspace{0.5cm}

As presented in \Cref{sec:results}, developers can identify architectural problems and reflect on solutions by visualising and analysing game engine architectural models. While useful for analysing game engines individually, we can also combine models from different game engines to observe which architectural patterns emerge from this combination, and which are, therefore, shared among all the game engines in the analysed set. 

\begin{figure}[ht]
    \centering
    \includegraphics[width=0.72\textwidth]{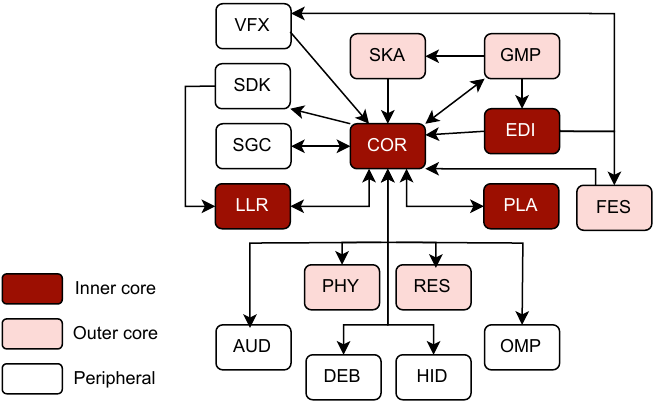}
    \caption{Our emergent open-source game engine architecture.}
    \label{fig:emergent_arch}
\end{figure}

For example, in \Cref{fig:emergent_arch}, we use a box-and-line diagram to represent the most frequent relationships between subsystems on the 10 game engines we analysed. In the centre of the diagram, we placed the subsystems with the highest betweenness centrality, forming an inner core (dark red). Next, we placed other subsystems which appear in \Cref{tab:results_frequent_coupling} in the outer core (light red). Finally, we placed the subsystems which do not appear in \Cref{tab:results_frequent_coupling} in the outer core's periphery (white). All relationships shown in the diagram are among the most frequent, as shown in \Cref{tab:results_frequent_coupling}. When there was a tie (e.g., two pairs had the same frequency), we chose the coupling pair with the highest sum of betweenness centrality. 

In this emergent architecture, we observe that \textit{Low-Level Renderer} (LLR) often inter-depends on \textit{Core} (COR), which it uses to access functionality in the \textit{Platform Compatibility Layer} (PLA) and \textit{Resources} (RES). It is often included by the \textit{World Editor} (EDI) and \textit{Gameplay Foundations} (GMP), which are both visual interfaces between the user and the game engine. Because it manages UI elements that emit events and trigger actions throughout the system, \textit{Front End} (FES) often depends on the event/messaging system in \textit{Core} (COR).

By providing a high-level view of subsystem \textit{includes}, an emergent game engine architecture such as we show here can provide an architectural reference to game engine developers wishing to build a new game engine that structurally resembles the set of game engines we analysed. It may also serve as a guide to impact analysis of an existing game engine, showing which subsystems might be affected by a change in any other number of subsystems.

\section{Controlled Experiment}
\label{sec:evaluation}

We conduct a controlled experiment with 16 developers to determine the qualitative success of SyDRA in supporting developers' understanding and maintenance of game engines. We base our study design on another similar study by \citeauthor{briand_controlled_2001} (\citeyear[p.~518]{briand_controlled_2001}), henceforth called ``original experiment''. In this section, we describe the design and execution of our controlled experiment, presenting and discussing the obtained results.

\begin{table}[ht]
\centering
\caption{Experimental Design.}
\label{tab:design}
\begin{tabular}{|c|cc|}
\hline
\textbf{\phantom{a}Variable X\phantom{a}} & \multicolumn{2}{c|}{\textbf{Variable Y - Tool}}    \\ \hline
\textbf{Run}        & \multicolumn{1}{c|}{\textbf{\phantom{a}VS Code\phantom{a}}} & \multicolumn{1}{c|}{\textbf{\phantom{a}Moose + VS Code\phantom{a}}} \\ \hline
1                   & \multicolumn{1}{c|}{A}            & \multicolumn{1}{c|}{B}              \\ \hline
\end{tabular}
\end{table}

We employed a between-group 2 x 1 design, as described in \Cref{tab:design}. The independent variables are the experimental runs (X) and the tools used to analyse Godot (Y). The assignment of participants to the control (A) and treatment (B) groups was done randomly to control learning and fatigue effects. Tasks were shown randomly to participants, except for Tasks 3 and 4, which depended on each other and could not be understood if shown in reverse order. We provide more details about our task choices in \Cref{sub:exp-tasks}.

\subsection{Hypotheses}
\label{sub:exp-hypoteses}

The null hypothesis is stated as:
\begin{itemize}
    \item \textit{H$_0$}: Using SyDRA provides no significant difference in the understandability and maintainability of game engine architecture.
\end{itemize}

\noindent
The alternative hypotheses, i.e., what is expected to occur, are stated as:
\begin{itemize}
    \item \textit{H$_1$}: It is significantly easier to understand game engine architecture \\ by using SyDRA.
    \item \textit{H$_2$}: It is significantly easier to perform impact analysis (locate changes) \\ in game engines by using SyDRA.
\end{itemize}

\subsection{Participants}
\label{sub:exp-participants}

We recruited 16 participants, all over 18 years of age and with prior experience in object-oriented programming. We recruited them via email or by asking them in person. Most participants are men under 30 based in Brazil or Canada. They are mostly students, researchers or software developers outside the video game industry. They have mostly 2 to 5 years of software development experience and have used Unity for student or hobby projects. 

Participants' familiarity with game engine usage and development varied greatly. For example, while 57\% of the participants reported no experience with game engines, two participants reported developing their own game engines. This diversity of levels of experience is important because it allows us to observe how the tools we selected for the study were used differently by each kind of developer and their challenges. A more detailed breakdown of demographics can be found in the replication package we published on Zenodo \footnote{\url{https://zenodo.org/records/11002050}}.

\subsection{Materials}
\label{sub:exp-materials}

In this study, participants analysed Godot\footnote{\url{https://github.com/godotengine/godot}}, a cross-platform, free and open-source game engine released by Juan Linietsky and Ariel Manzur in 2014. We chose Godot due to its relevance to the open-source developer community on GitHub, as explained in \Cref{sec:approach}. While control group participants used exclusively Visual Studio Code to analyse Godot's source code, treatment group participants used both Moose + Visual Studio Code, and therefore had access to the Architectural Map visualisation of Godot produced with SyDRA.

We created instructional documents to teach participants to use the given tools. As we explain in \Cref{sub:exp-procedures}, we asked participants to read the document related to the tool they were about to use before performing the tasks. These documents provided an overview of features such as file searching and browsing, illustrated by screenshots. At the end of each document, we also provided three optional warm-up exercises designed to reinforce the instructions. Although similar to the tasks in the experiment, these warm-ups were simpler and focused on a single feature at a time. For example, in the Visual Studio Code warm-ups, we asked participants to use the search feature to find and count the number of files containing the word ``Music''. They could then verify their answers on the last page of the document.

In this section, we provide an overview of the main features of Moose and Visual Studio Code and our rationale for choosing these tools for the study. We also describe the features participants used during the experiment to aid them in the completion of the tasks and show examples of the screenshots included in the instructional documents.

\vspace{0.3cm}
\noindent
\textbf{Moose}: A platform for software analysis\footnote{\url{https://moosetechnology.org/}} composed of several tools built on top of the Pharo\footnote{\url{https://pharo.org/}} programming language. It enables users to define metamodels and create models based on them. It also allows users to visualise these models as Architectural Maps, as explained in \Cref{sec:approach}. Model entities can be written (or ``propagated'' in Moose's jargon) to a bus, which is a channel of communication between tools \cite[p.130]{ben_sassi_modular_2020}. Moose tools, such as the Architectural Map, can then read entities from the bus and do something with them (e.g., draw a visualisation).

\begin{figure*}[ht]
   \centering
   \includegraphics[width=0.75\linewidth]{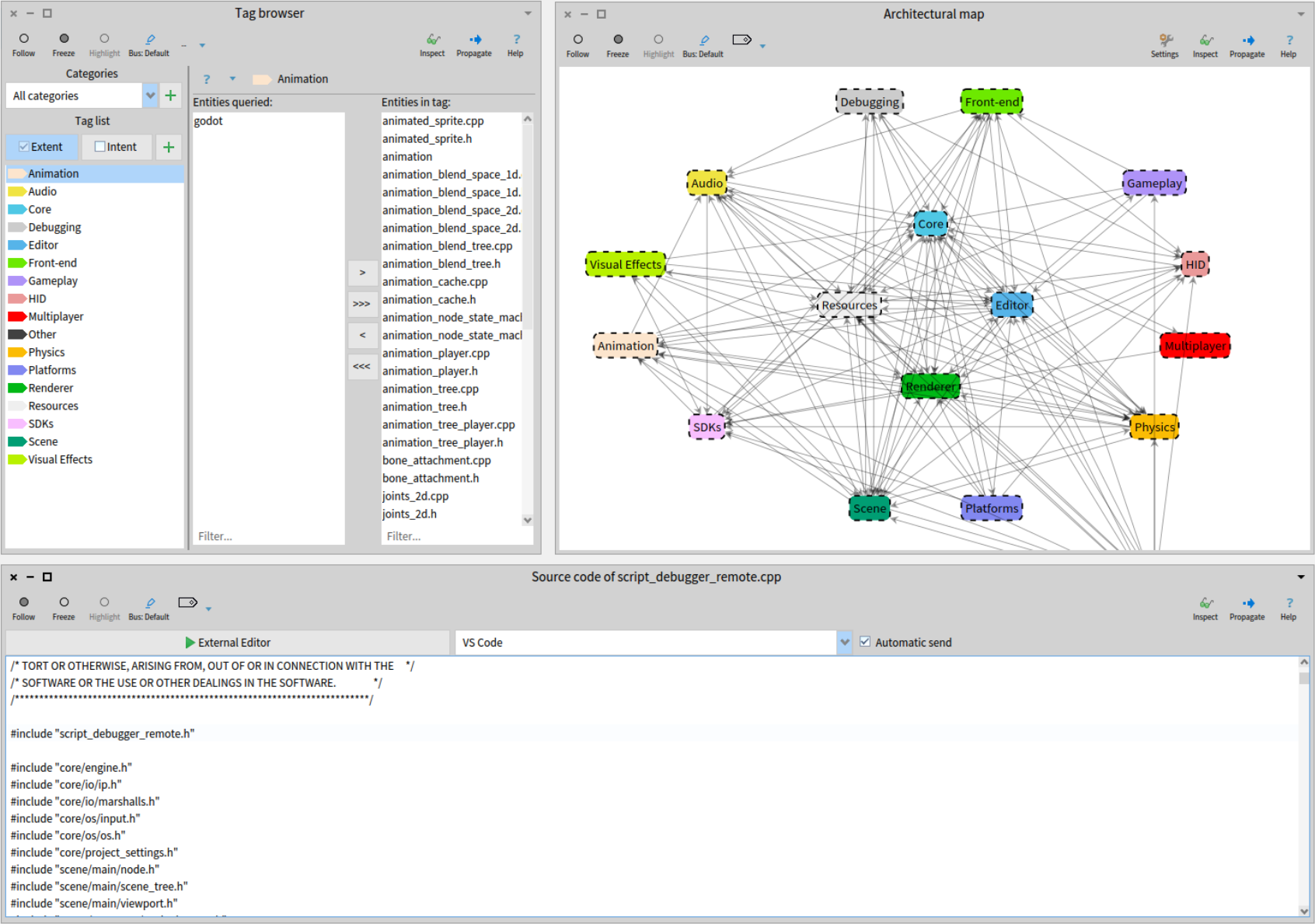}
    \caption{Moose with Architectural Map visible on the top right.}
    \label{fig:moose-exp}
\end{figure*}

During the study, as shown in \Cref{fig:moose-exp}, treatment group participants located files and folders using the Architectural Map (top right) and propagated them to a built-in source code browser to inspect the source code (bottom). Participants could launch Visual Studio Code from Moose built-in editor to use features such as code folding, syntax highlighting, and search, not available on Moose. They could also use Moose tag browser to see the list of files clustered into each subsystem, represented as coloured tags (top left).

\begin{figure*}[ht]
    \centering
    \includegraphics[width=0.9\linewidth]{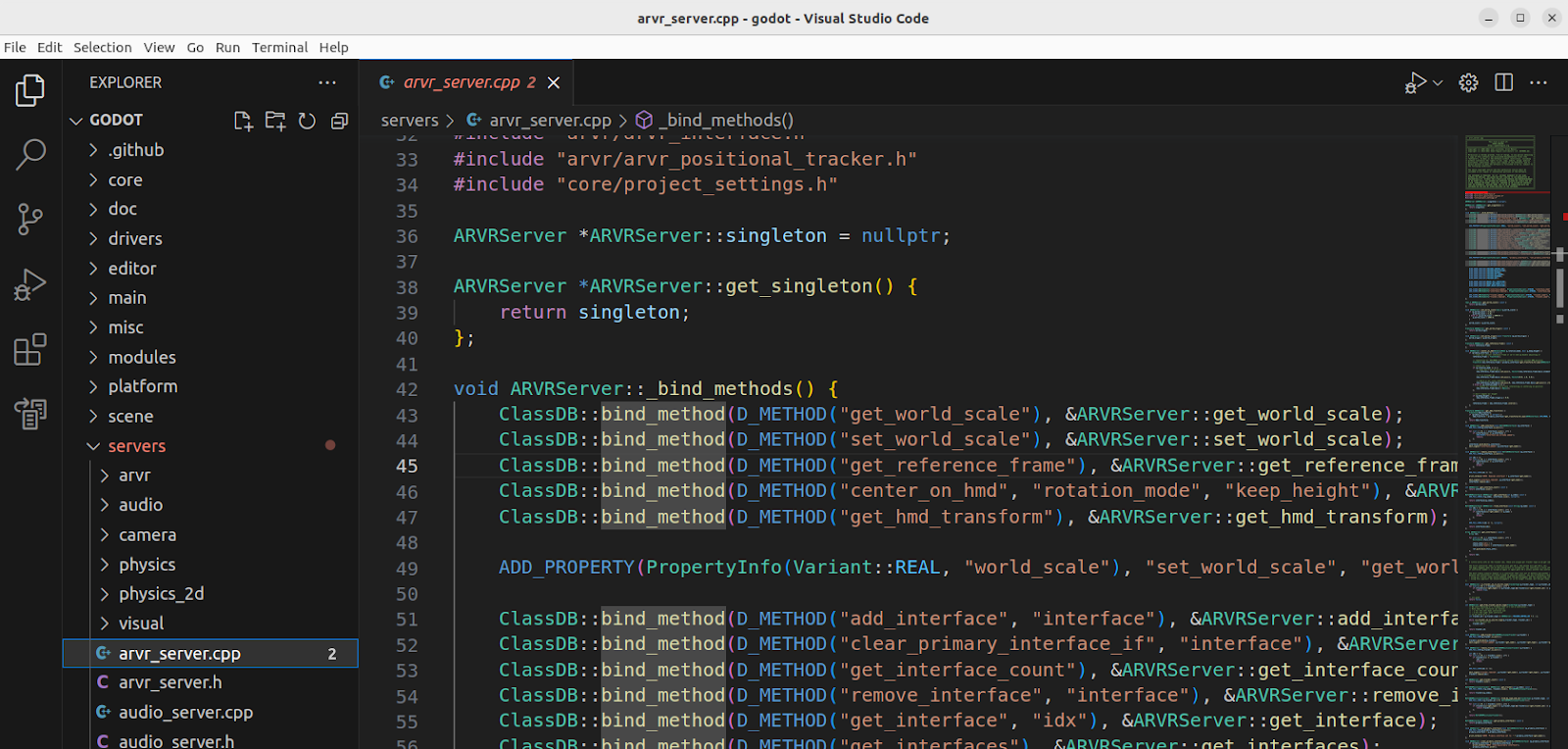}
    \caption{Visual Studio Code, used by the control group participants.}
    \label{fig:vscode-exp}
\end{figure*}

\vspace{0.3cm}
\noindent
\textbf{Visual Studio Code}: A source code editor\footnote{\url{https://code.visualstudio.com/}} released by Microsoft in 2015. Also commonly referred to as VS Code. During the study, control group participants used it to locate files and folders inside the Godot repository, read source code and search for words as directed by the task statements, as we show in \Cref{fig:vscode-exp}. We chose VS Code due to its broad popularity among software developers. According to the Stack Overflow Developer Survey 2023, 74.09\% of professional developers and 78.39\% of developers learning to code use VS Code\footnote{\url{https://survey.stackoverflow.co/2023/\#technology-most-popular-technologies}}.

\vspace{-0.5cm}
\subsection{Tasks}
\label{sub:exp-tasks}

We asked participants to perform nine tasks during the study, which were of two kinds: architectural understanding and impact analysis. In architectural understanding tasks, we asked participants to explain game engine subsystems and dependencies between files. In impact analysis tasks, we asked participants to point out which files should be changed due to a change/removal of functionality in another part of the system. Participants performed seven architectural understanding tasks and two impact analysis tasks.

\begin{table}[ht]
\centering
\caption{Examples of task statements}
\label{tab:example-tasks}
\begin{tabular}{@{}p{2.5cm}p{4.5cm}p{4.5cm}@{}}
\toprule
\textbf{Type} & \textbf{Control Group} & \textbf{Treatment Group} \\ 
\midrule
Architectural \newline Understanding & Search for the file servers/audio/audio\_effect .h and open it. Provide a short description of its functionality.                                           & Expand the Audio subsystem, expand the “audio” folder and propagate the file servers/audio/audio\_effect.h. Provide a short description of its functionality. \\
Impact \newline Analysis & Suppose the rich text functionality in the Front end subsystem of Godot was removed. Please mention all files which may have to be changed as a result of the removal of these functionalities. & Suppose the rich text functionality in the Front end subsystem of Godot was removed. Please mention all files which may have to be changed as a result of the removal of these functionalities. \\
\bottomrule
\end{tabular}
\end{table}

We wrote our task statements based on those provided in Appendix A of the original experiment \cite[p.~527]{briand_controlled_2001}. We changed the statements slightly to make them easier for novice developers to understand. We also adapted task statements to reflect the steps participants in different groups had to perform to find files in the tools they were using, as demonstrated in \Cref{tab:example-tasks}. For example, while in Task 2, we asked treatment group participants to ``Expand the Audio subsystem'' and then ``propagate'' a given file, we asked control group participants to search the file by name and then open it\footnote{See Folder 2 of the replication package for the complete task statements.}. This way, we ensured both participants were directed to the same file, even though they followed different steps to find and open it.

\subsection{Procedures}
\label{sub:exp-procedures}

The study session was divided into three parts. First, we asked participants to read and follow the instructional document described in \Cref{sub:exp-materials}. Then, we asked them to perform the tasks. Finally, participants completed a debriefing questionnaire, where we asked them for background information and also for a workload assessment of the tasks they performed. For the second and third parts, participants submitted their answers via an online form, which also computed the time elapsed between answers.

The debriefing questionnaire was divided into two parts\footnote{See Folder 2 of the replication package for the complete debriefing questions/answers.}. In questions 1 to 7, we asked participants about their professional backgrounds and demographics. In questions 8 to 13, we asked participants to make a workload assessment of the tasks they performed based on the NASA TLX (Task Load Index) questionnaire. We chose NASA TLX because it has been used for over 20 years by several studies that evaluate software development \cite[p.668]{al_madi_assessing_2022}, interface design and decision-making activities \cite[p.906]{hart_nasa-task_2006}.

We used the information provided by participants in the debriefing questionnaire to qualitatively measure their perception of effort and stress concerning the tools they used and the tasks they performed. We also correlated their performance with their years of development experience and familiarity with game engines to understand how each variable influences performance. We discuss these comparisons in more detail in \Cref{sec:results}.

We remained available throughout the study session to support participants but kept physically distanced from them. Our support was limited to clarifying task descriptions when asked and resolving technical issues with the computer and tools related to the study.

\subsection{Measurements}
\label{sub:exp-design}
We measured the level of game engine architectural understanding by the participants by measuring the time they spent on tasks and how correctly they completed tasks. We derived six dependent variables from this data, as in the original experiment \cite[p.~518]{briand_controlled_2001}:

\begin{itemize}
    \item \textit{UndTime}: Time spent on architectural understanding tasks in minutes.
    \item \textit{UndCorr}: Correctness of architectural understanding tasks (e.g., the number of tasks correctly answered). 
    \item \textit{ModTime}: Time spent on impact analysis tasks in minutes.
    \item \textit{ModComp}: Completeness of the impact analysis, obtained by dividing the number of correct files informed by the participant by the actual number of correct files. 
    \item \textit{ModCorr}: Correctness of the impact analysis, obtained by dividing the number of correct files informed by the participant by the total number of files informed by the participant. 
    \item \textit{ModRate}: Modification rate, obtained by dividing the number of correct files informed by the participant by \textit{ModTime}.
\end{itemize}

The original study did not define architectural understanding correctness, so we defined it as binary: an answer is either correct (1) or incorrect (0). Therefore, \textit{UndCorr} ranges from zero to seven, given there were seven architectural understanding tasks in total. For Impact Analysis tasks, correctness is defined as the ratio between the number of files the participant informed correctly, and the total number of files they informed. Therefore, \textit{ModCorr} ranges from zero to two, given there were two impact analysis tasks in total.

\subsection{Data Analysis Procedures}
\label{sub:exp-analysis}

We collected data from participants throughout 16 sessions spread across one month. Therefore, eight data points were available for the control group and eight for the treatment group. All participants answered all tasks and debriefing questions. We used the following statistical techniques\footnote{See Folder 3 of the replication package for statistical testing results.} to determine whether the data collected during the study was statistically significant:

\begin{enumerate}
    \item \textbf{Number of participants:} We used a two-sample T-test to determine the number of participants we would need to detect statistically significant differences. We considered $\alpha=0.05$, a statistical power of 0.9 and standard deviations based on the original experiment. For task completion time, we defined a minimum difference of 1 minute. For task correctness, we defined a minimum 30\% difference. The minimum number of participants would be 14 in total. We exceeded this number by conducting the study with 16 participants to ensure statistical significance.
    \item \textbf{Normality:} We used both the Kolmogorov-Smirnov and the Shapiro-Wilks' W normality tests. In both cases, we observed all dependent variables had non-normal distributions.
    \item \textbf{Statistical Significance:} We used a non-parametric significance test that is adequate for non-normal data, the Wilcoxon Matched Pairs test. To be significant, the result of the test, called the Z value, must exceed the critical Z value for $\alpha=0.05$, one-tailed, as provided by the Wilcoxon Signed-Ranks Table\footnote{\url{https://real-statistics.com/statistics-tables/wilcoxon-signed-ranks-table/}}. We observed \textit{UndTime}, \textit{UndCorr}, \textit{ModCorr} and \textit{ModRate} exceeded the critical Z value, while \textit{ModTime} and \textit{ModComp} did not. 
    \item \textbf{Effect Size:} We computed the effect size for each variable by calculating the difference between the control group and treatment group arithmetic means, then dividing the result by the geometric mean of the control group and treatment group standard deviations.
\end{enumerate}

\subsection{Results}
\label{sub:exp-results}

\begin{figure}[h]
    \centering
    \begin{subfigure}{0.24\linewidth}
        \includegraphics[width=\linewidth]{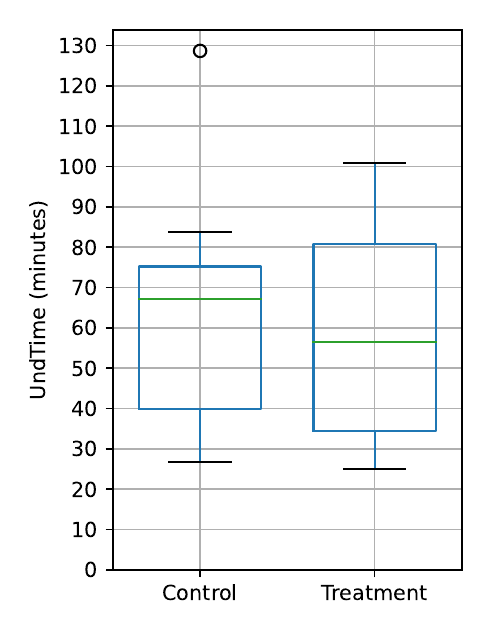}
        \caption{\textit{UndTime}}
        \label{fig:results_undtime}
    \end{subfigure}
    \hfill
    \centering
    \begin{subfigure}{0.24\linewidth}
        \includegraphics[width=\linewidth]{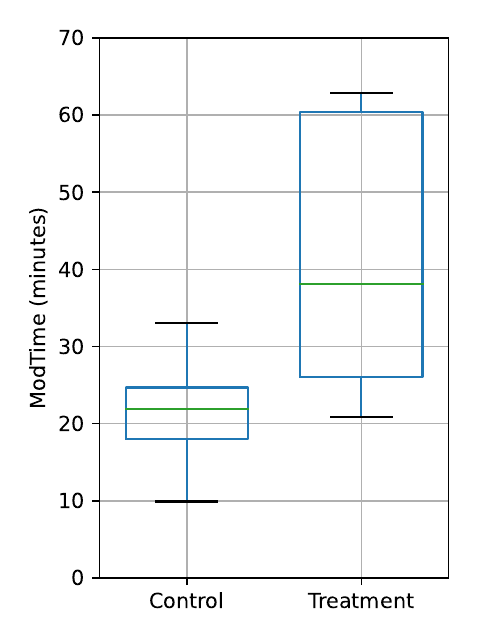}
        \caption{\textit{ModTime}}
        \label{fig:results_modtime}
    \end{subfigure}
    \centering
    \begin{subfigure}{0.24\linewidth}
        \includegraphics[width=\linewidth]{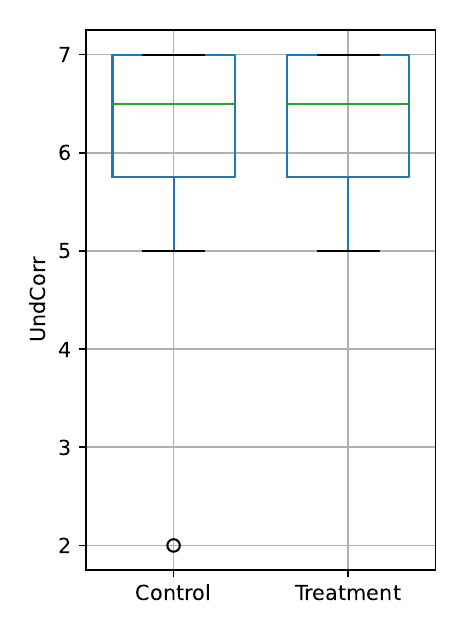}
        \caption{\textit{UndCorr}}
        \label{fig:results_undcorr}
    \end{subfigure}
    \hfill
    \centering
    \begin{subfigure}{0.24\linewidth}
        \includegraphics[width=\linewidth]{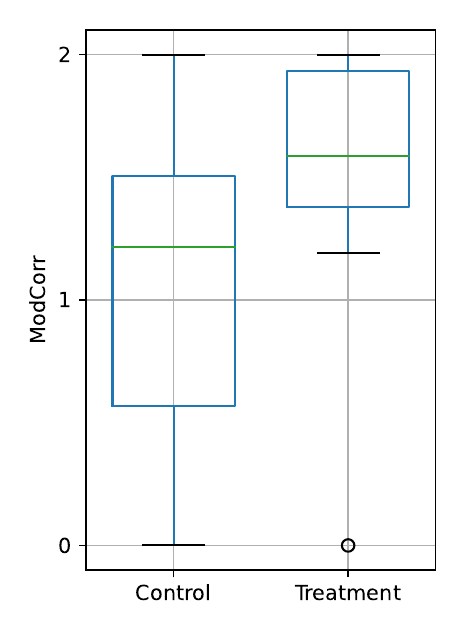}
        \caption{\textit{ModCorr}}
        \label{fig:results_modcorr}
    \end{subfigure}
    \caption{Correctness and completion time distribution for both study groups}
    \label{fig:results-boxplots}
\end{figure}

In \Cref{tab:results-descriptive} we show a summary of the dependent variables collected from the 16 participants of the study. The columns represent the mean ($\overline{X}$), the median ($\tilde{m}$), minimum and maximum values and standard deviation (s). On average, participants took 62 minutes to complete understanding tasks and 31 minutes to complete impact analysis tasks, totalling 1 hour and 33 minutes. From the standard deviation, we observe a high variability in both completion time and correctness, reflecting the participants' diverse levels of experience. 
\vspace{-0.3cm}
\begin{table}[ht]
    \centering
    \caption{Descriptive statistics for each dependent variable.}
    \label{tab:results-descriptive}
    \begin{tabular}{lrrrrrl}
        \toprule
        \textbf{Variable} & \textbf{\phantom{a}$\overline{X}$} & \textbf{\phantom{aaa}$\tilde{m}$} & \textbf{\phantom{a}min} & \textbf{\phantom{a}max} & \textbf{\phantom{aaa}s} & \textbf{measurement} \\ \midrule
        \textit{UndTime}           & 62.0                    & 67.3                & 25.0        & 128.7        & 30.0 & minutes \\
        \textit{UndCorr}           & 6.1                     & 6.5                 & 2.0        & 7.0         & 1.3   & tasks    \\
        \textit{ModTime}           & 31.4                    & 25.4                & 10.0         & 62.9        & 17.1 & minutes   \\
        \textit{ModCorr}           & 1.3                     & 1.4                 & 0.0         & 2.0         & 0.7  & ratio actual/expected correct file count     \\
        \textit{ModComp}           & 1.5                     & 1.1                 & 0.6         & 3.6         & 0.9  & ratio provided correct/total file count     \\
        \textit{ModRate}           & 0.0                     & 0.0                 & 0.0         & 0.0         & 0.0  & ratio correct files/\textit{ModTime}    \\ 
        \bottomrule
    \end{tabular}
\end{table}
\clearpage
By observing variables related to time (\textit{UndTime}) and correctness (\textit{UndCorr}) together in \Cref{fig:results_undtime} and \Cref{fig:results_undcorr}, we can understand how the tool used by each group influenced the performance of participants in the architectural understanding tasks. While the treatment group completed the tasks faster, both groups completed them with the same level of correctness. Therefore, using SyDRA decreases task completion time but has no effect on task correctness.
\vspace{-0.3cm}

\vspace{-0.5cm}
In impact analysis tasks, the control group completed tasks faster but also less correctly than the treatment group, as we show in \Cref{fig:results_modtime} and \Cref{fig:results_modcorr}. This happened because, by using exclusively VS Code, participants had more difficulty finding all outgoing and incoming \textit{include} relationships between files and ended their analysis prematurely. In contrast, participants using SyDRA could succinctly visualize relationships as arrows in the Architectural Map, which helped them complete impact analysis more correctly.
\vspace{-0.6cm}
\begin{table}[ht]
\centering
\caption{Effect size for each dependent variable}
\label{tab:results-effect-size}
\begin{tabular}{lrrrr}
\toprule
\textbf{Variable} & \textbf{\phantom{a}$\overline{X}$ Control} & \textbf{\phantom{a}$\overline{X}$ Treatment} & \textbf{\phantom{a}Effect Size} & \textbf{\phantom{aaa}Z} \\
\midrule
\textit{UndTime}    & 65.5   & 49.2  & 0.4   & 20.0    \\
\textit{UndCorr}    & 5.9    & 6.2   & 0.3   & 10.5    \\
\textit{ModTime}    & 21.4   & 32.0  & 1.7   & 5.0     \\
\textit{ModComp}    & 1.0    & 2.0   & 1.5    & 5.0      \\
\textit{ModCorr}    & 1.1    & 1.5   & 0.5    & 7.0      \\
\textit{ModRate}    & 0.0    & 0.0   & 0.2    & 12.0     \\
\bottomrule
\end{tabular}
\end{table}

\vspace{-0.3cm}
However, we must also consider statistical significance and effect size when interpreting these results. For example, while \textit{UndTime}, \textit{UndCorr} and \textit{ModCorr} are statistically significant, \textit{ModTime} is not, which means the average impact analysis time observed in this study cannot be generalised. Moreover, we did not identify a large effect size for any statistically significant variables, as shown in \Cref{tab:results-effect-size}. Considering the scale defined by \cite{kampenes_systematic_2007} for software engineering, the largest effect size for a statistically significant variable was ``medium'' for \textit{ModCorr}.

\vspace{.5cm}
\noindent
\fbox{
    \parbox{0.95\textwidth}{
        \textbf{Conclusion of the Controlled Experiment}: \newline
        Based on our observations, we accept both \textit{H$_1$} and \textit{H$_2$}. The results show that using SyDRA enables a statistically significant decrease in task completion time while not affecting task correctness. Regarding impact analysis, using SyDRA results in slightly higher task correctness but no statistically significant difference in task completion time.
    }\vspace{1cm}
}

\subsection{Discussion}
\label{sub:exp-discussion}

\begin{figure}[ht]
	\begin{subfigure}{0.25\linewidth}
		\includegraphics[width=\linewidth]{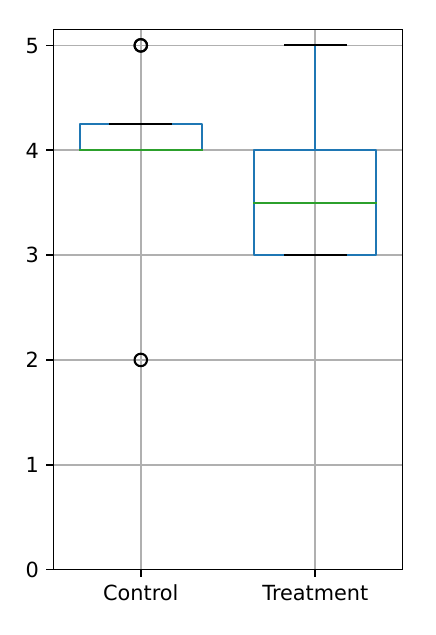}
		\caption{Mental}
		\label{fig:results_a1_mental}
	\end{subfigure}
	\hfill
	\begin{subfigure}{0.25\linewidth}
		\includegraphics[width=\linewidth]{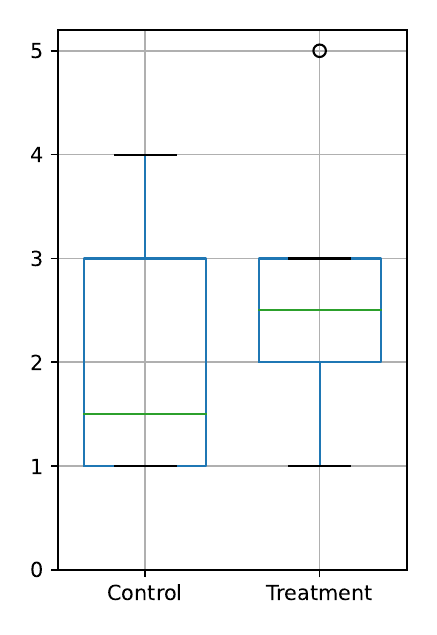}
		\caption{Physical}
		\label{fig:results_a2_physical}
	\end{subfigure}
	\hfill
	\begin{subfigure}{0.25\linewidth}
		\includegraphics[width=\linewidth]{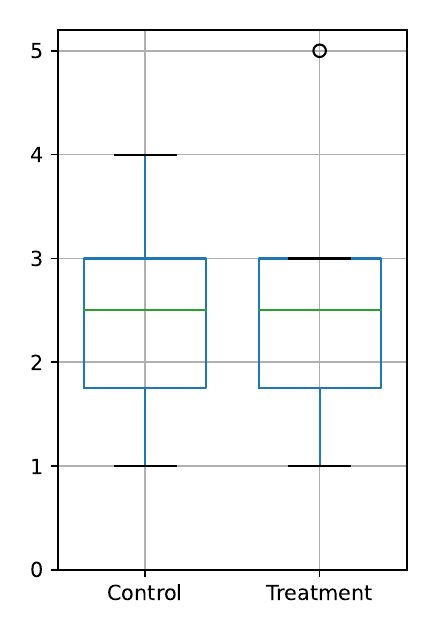}
		\caption{Temporal}
		\label{fig:results_a3_temporal}
	\end{subfigure}
	\hfill
	\begin{subfigure}{0.25\linewidth}    
		\includegraphics[width=\linewidth]{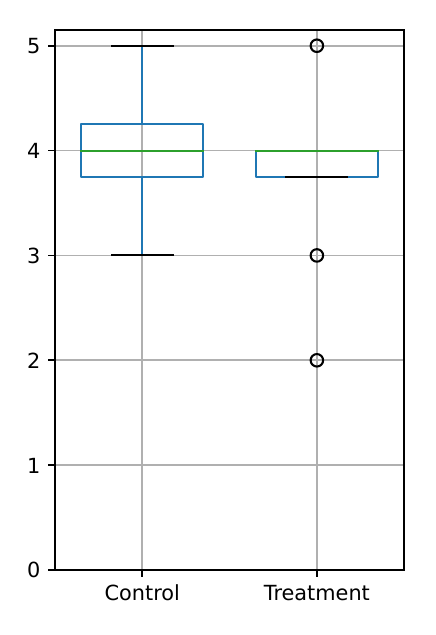}
		\caption{Success}
		\label{fig:results_a4_success}
	\end{subfigure}
	\hfill
	\begin{subfigure}{0.25\linewidth}
		\includegraphics[width=\linewidth]{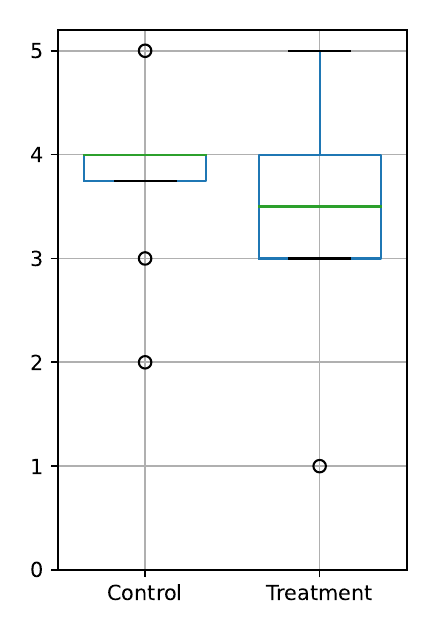}
		\caption{Effort}
		\label{fig:results_a5_effort}
	\end{subfigure}
	\hfill
	\begin{subfigure}{0.25\linewidth}
		\includegraphics[width=\linewidth]{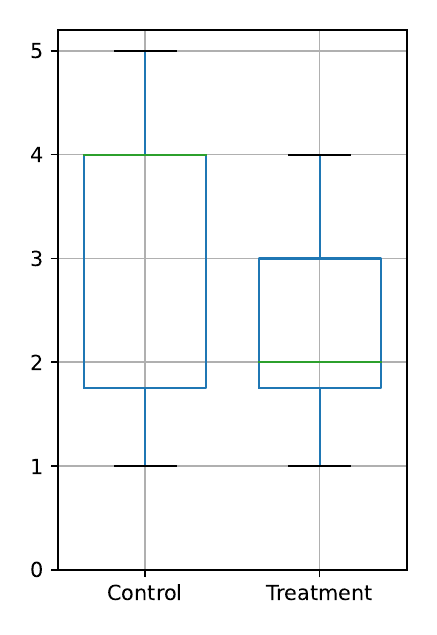}
		\caption{Frustration}
		\label{fig:results_a6_frustration}
	\end{subfigure}
	\caption{Participants' answers for NASA TLX questions}
	\label{fig:results_debriefing_tlx}
\end{figure}

In \Cref{fig:results_debriefing_tlx}, we compare participant answers about six aspects of task load described in the NASA TLX questionnaire: mental, physical and temporal demand, perception of success, effort and frustration. As for the perception of success and temporal demand, there was no difference between groups, which is evidence that the tools the participants used did not make them feel overwhelmed. 

\clearpage

We also observe that the participant's perception of success correlates with their experience level. For example, video game developers reported a higher perception of success, lower mental demand and lower frustration than non-video game developers and students/researchers. We observe the same pattern when comparing novice (less than five years of professional experience) and experienced (five years of professional experience or more) developers of all backgrounds.

The treatment group reported lower mental demand, perception of effort and frustration when compared to the control group, which is evidence that participants felt more comfortable using VS Code with the SyDRA-generated architectural model instead of VS Code only. In contrast, the treatment group reported higher physical demand than the control group. However, physical demand is not a major contributor to workload in software development \cite[p.671]{al_madi_assessing_2022}. Therefore, we believe these reports of high physical demand are not evidence that using SyDRA necessarily demands more body movement or effort than VS Code only. 

Overall, the data we collected through the NASA TLX questionnaire shows that participants perceive a lower task load when using SyDRA, even though this perception correlates to each participant's professional experience and familiarity with the video game domain. The largest difference we observed between the groups was in the perception of task mental demand and frustration.

\section{Threats to validity}
\label{sec:threats}

In this section, we discuss threats to the validity of SyDRA and the controlled experiment.

\subsection{External Validity}
\label{sub:threats-external}
We acknowledge that the game engines we selected for analysis with SyDRA may not be entirely representative of all open-source game engines or the entire video game industry. We mitigated this issue by selecting game engines based on their popularity, as described in \Cref{sec:approach}. We confined our analysis to C++ game engines, which may have led to the exclusion of pertinent game engines developed in other programming languages. 

Moreover, we acknowledge using ``Runtime Game Engine Architecture'' \cite[p.~33]{gregory_game_2018} in subsystem detection across all game engines, potentially introducing a bias. As a mitigation strategy, in future work, we intend to encompass a wider spectrum of subsystems, both obtained via SyDRA and game engine development literature.

In the controlled experiment, we used a 2x1 design, not within-subject, which differs from the original 2x2 within-subject design. We chose this design due to limited participant availability, which made it hard or sometimes impossible to ensure all participants could participate twice in the study. As explained in the original experiment, by using the 2x2 within-subject design ``the error variance due to differences among subjects is reduced'' \cite[p.518]{briand_controlled_2001}. We are aware that by choosing the 2x1 design we risked obtaining higher error variance for all dependent variables. In future work, we intend to run another study with more participants and also use a 2x2 within-subject design.

Moreover, a confounding effect may result from our selection of game engines, analysis tools and participants for the controlled experiment. For example, Godot may not be representative of all open-source game engines in size and complexity. Also, most participants did not have prior experience with video game development and therefore do not accurately represent developers in these domains. Finally, we did not detect a large effect size for any of the statistically significant dependent variables, which is evidence that our results may not generalize to other game engines or more diverse participants.

\subsection{Internal Validity}
\label{sub:threats-internal}
The subsystem detection step of SyDRA was performed manually by the first author, which may have introduced a bias in the process. To mitigate this issue, we intend to assign multiple people to work in this step and later combine their results by consensus. We also intend to explore quasi-automated approaches for subsystem detection to determine the most suitable method for game engines and other types of software.

Moreover, we are aware that our analysis of SyDRA's results is dependent on Moose behaviour, and also on the graph metric (e.g., in-degree and centrality) we extracted from the architectural models. Changing them could also change the results and therefore our perception of these game engine architectures. In future work, we intend to experiment with different software analysis and visualisation tools and measure to what extent they can help developers perform architectural understanding, impact analysis and testing activities.

In the context of the controlled experiment, while we measured how quickly and correctly participants completed understanding and impact analysis tasks, we did not measure whether they would be able to implement the changes in the source code quickly and correctly as well. Also, while tasks allowed for several possible solutions, we did not verify whether the solution provided by the participant was the best or most optimized in practice, only whether it was architecturally sound.

We are aware that most task statements and debriefing questions allowed for multiple interpretations and that may have been the reason for the large variation in terms of task completion time and task correctness we observed. An observer effect may have also contributed to this variation, considering participants may have felt less stressed or behaved differently if they were not being observed. As we explained in \Cref{sub:exp-procedures}, we tried to mitigate this effect by physically distancing from the participants and interacting with them only when necessary.

A maturation effect may have occurred in the study due to participants learning how the tools work as the study proceeded. Some participants also had prior experience with the tools used in the study, which might have helped them complete tasks faster and more correctly than others. As stated in \Cref{sub:exp-materials}, we mitigated this effect by choosing a tool that is known by most developers (Visual Studio Code), as well as providing instructional documents for both tools.

An instrumentation effect may also have occurred due to differences between control and treatment task descriptions. As explained in \Cref{sub:exp-tasks}, we did our best to ensure the tasks could be completed with a similar amount of effort in both tools and that the task descriptions were clear and stated in the same way to both groups. However, it is possible that these differences also influenced the participants' abilities to understand and complete tasks more correctly.


\section{Conclusion}
\label{sec:conclusion}
In this paper, we present the Subsystem-Dependency Recovery Approach (SyDRA). By applying this approach to 10 open-source game engines, we obtain architectural models which can be used to compare game engine architectures and identify and solve problems such as high coupling and low cohesion. Additionally, we showed and discussed ways in which the metrics and visualisations derived from architectural models can aid game engine development, such as:

\begin{itemize}
    \item \textbf{Architectural Understanding}: Architectural model visualisations provide a friendly way for novice game engine developers to understand this kind of system and start developing their own subsystems or plugins. Moreover, we show how the use of architectural models can aid architectural understanding in \Cref{sub:exp-results}.
    \item \textbf{Impact Analysis}: Game engine developers can refactor their code more safely by visualising how changes to a subsystem could impact the whole game engine. 
    \item \textbf{Reference Extraction}: Game engine architects seeking to design a new engine can extract architectural models from similar systems and use them as references. They can either extract and visualize data for a single system or join data from multiple systems within the same family, as we show in \Cref{fig:emergent_arch}. This is useful both for large companies and small indie developers who develop tailor-made solutions, e.g., for performance. 
\end{itemize}

Through a controlled experiment with 16 software developers, we evaluated the extent to which SyDRA helps developers perform architecture understanding and impact analysis tasks. We asked developers to analyse Godot, an open-source game engine. While control group participants used exclusively VS Code to analyse Godot's source code, treatment group participants used both Moose + Visual Studio Code, and therefore had access to the Architectural Map visualisation of Godot produced with SyDRA. Our experiment's most important contribution is showing a small yet statistically significant decrease in completion time for architectural understanding tasks when using VS Code in conjunction with the SyDRA-generated architectural model. 

Finally, our analysis of the answers in the experiment's debriefing questionnaires shows that by using SyDRA developers can better understand game engine architecture while not increasing their perceived workload. In future work, we intend to ask game engine developers to evaluate our suggestions for subsystem coupling reduction (\Cref{sec:subsystem-aspect}) and subsystem cohesion increase (as shown in \Cref{sec:folder-level-ref}), as well as the emergent game engine architecture we propose in \Cref{sec:discussion}. We aim to understand whether these suggestions are useful in a professional video game development setting, and we will seek collaboration with independent video game development companies to achieve this.

Moreover, we are aware SyDRA is flexible enough to be used in the analysis of other software families which are fundamental to video game production, such as image and sound editors and 3D modelling software. Therefore, we intend to apply SyDRA to these software families, to identify common architectural structures that can guide their creation, maintenance and evolution.

\section*{Acknowledgements}
The authors were partially supported by the NSERC Discovery Grant and Canada Research Chairs programs.

\bibliographystyle{ACM-Reference-Format}
\bibliography{main.bib}

\end{document}